# A Deep Learning Approach to Radar-based QPE

Enter authors here: Ting-Shuo Yo[1,4], Shih-Hao Su[2], Jung-Lien Chu[3], Chiao-Wei Chang[2], and Hung-Chi Kuo[1]

[1]National Taiwan University.

[2]Chinese Culture University.

[3]National Science and Technology Center for Disaster Reduction.

[4]DataQualia Lab. Co. Ltd.

Corresponding author: Ting-Shuo Yo (tsyo@ntu.edu.tw)

†Additional author notes should be indicated with symbols (current addresses, for example).

**Key Points:**

- We proposed a novel deep learning approach for estimating precipitation from a large data volume and demonstrated it with QPE from aggregated radar data

- In comparison to the operational QPE scheme, the proposed framework performed comparably well in general cases and excelled in detecting heavy-rainfall events

- The proposed framework can be further extended to integrate heterogeneous data sources and potentially improve the QPF in extreme precipitation scenarios

**Abstract**

In this study, we propose a volume-to-point framework for quantitative precipitation estimation (QPE) based on the QPESUMS (Quantitative Precipitation Estimation and Segregation Using Multiple Sensor) Mosaic Radar dataset. With a data volume consisting of the time series of gridded radar reflectivities over the Taiwan area, we used machine learning algorithms to establish a statistical model for QPE in weather stations. The model extracts spatial and temporal features from the input data volume and then associates these features with the location-specific precipitations. In contrast to QPE methods based on the Z-R relation, we leverage the machine learning algorithms to automatically detect the evolution and movement of weather systems and associate these patterns to a location with specific topographic attributes. Specifically, we evaluated this framework with the hourly precipitation data of 45 weather stations in Taipei during 2013 ~ 2016. In comparison to the operational QPE scheme used by the Central Weather Bureau (CWB), the volume-to-point framework performed comparably well in general cases and excelled in detecting heavy-rainfall events. By using the current results as the reference benchmark, the proposed method can integrate the heterogeneous data sources and potentially improve the forecast in extreme precipitation scenarios.

**Plain Language Summary**

Quantitative Precipitation Estimation (QPE) is a method of approximating the amount of rain that has fallen at a location or across a region. In most use cases, weather service providers use radar signals to estimate the amount of precipitation through the formula describing the relationship between radar reflectivity and the size of raindrop particles, Z-R relation. The state-of-the-art QPE methods with adjusted Z-R relation is robust and accurate in general. Yet, they consider only the radar signal at a given location, which represented a point-to-point framework. In this study, we proposed a volume-to-point alternative that estimates the amount of rain by considering the signals in a broader spatial region and a longer time-span. By using a deep neural network to process the large data volume, we demonstrated that the proposed method performed comparably well in general cases and excelled in detecting heavy-rainfall events. This method could improve advanced warning for flash flooding and make water resource management more effective. In on-going research, we seek to extend our approach to integrate the heterogeneous data sources, and to be applied to precipitation forecasting.

**1 Introduction**

Heavy rainfall is one of the major causes of natural disasters on the mountainous island of Taiwan (Chang, 1996; Teng et al., 2006). Among all modern instruments for weather observations, the precipitation derived from the rain gauges and the weather radar is frequently compared against each other. As a common part of the standard weather service operation, rain gauges directly measure the amount of liquid precipitation falling to the ground. The measurement contains certain bias in nature (Collier, 1986; Yang et al., 1998; Habib et al., 1999; Habib et al., 2001; Ciach, 2002; Seo, 2002; Sevruk, 2005; Martinaitis et al., 2015), and the spatial density and locations of rain gauges are limited. Alternatively, weather radar provides Quantitative Precipitation Estimation (QPE) over a larger area with high spatial resolution, but it also incurs errors in the estimation process. Currently, most QPE is based on the Z-R relation developed by Marshall and Palmer (1948). Many studies had proposed adjustments to the precipitation in addition to the values derived from the Z-R relationship. Bellon and Austin (1984, 1987) analyzed the errors of QPE and Quantitative Precipitation Forecasting (QPF) and proposed adjustment schemes based on the growth and the movement of nearby storms. Steiner et al. (1999) improved the QPE results by 10% in terms of the root-mean-square-errors (RMSE) by using high‐quality gauge data and storm‐based bias adjustment. Chen et al. (2010) showed that the bias adjustment might come from topographic information of the rain gauge as well. Alternatively, Chiang et al. (2007) used Artificial Neural Network (ANN) to adjust the parameters in the Z-R relation and improved the QPE results. Besides improving the formula of the Z-R relationship, advancements in radar technology also improved the performance of QPE. Multi-radar networks have been shown to enhance the QPE results with composite reflectivity data (Crosson et al., 1996; Gourley et al., 2002; Zhang et al., 2011; Wu et al., 2012). Also, new generation radar systems such as S-band polarimetric radar can provide extra information and hence improve the QPE results (Jou et al., 2018).

The Z-R relationship originated from a point-to-point mapping between radar reflectivity and raindrop size distribution (Marshall & Palmer, 1948). The form of this relation, $Z = aR^b$, is robust and can be efficiently applied to the whole domain, but using the same set of formula coefficients ($a$ and $b$) for all locations or weather types could cause errors. Among the adjustments proposed by later studies, some used location-specific parameters (Chiang et al., 2007; Chen et al., 2010), and others added corrections based on extra information. For example, the evolution and movement of weather systems were significant modifiers (Bellon & Austin, 1984; Austin, 1987; Steiner et al., 1999), and using multiple radars or new radar designs were also helpful (Crosson et al., 1996; Gourley et al., 2002; Zhang et al., 2011; Wu et al., 2012; Jou et al., 2018).

In this study, we proposed a volume-to-point (VTP) approach to estimate precipitation from radar reflectivity in the Taipei Basin. There are 6 million people, about 1/3 of the Taiwan population, who reside in the Taipei Basin. The hill surrounding the Taipei basin can contribute to the heavy rain event in the metropolitan area of Taipei in the summer (Kuo & Wu, 2019). Most radar products cover a large spatial area that contains the information of weather systems such as convection patterns and convective cells. The merger of convective cells can produce heavy precipitation of more than 100 mm/hr in the Taipei basin (Miao & Yang, 2020). The data volume can span through both time and space. Using data in a broader spatial region is similar to the concept of the box-based spatial feature construction proposed by Han et al. (Han et al., 2017). And if we aggregate successive radar images over some time, the resulting data volume contains both spatial and temporal information. With proper analysis methods, we hope to extract the features concerning the evolution and movement of convective cells with adequate analyses. Therefore, the basic concept of using data volume is similar to that of Steiner et al. (1999). We also want to note that in this study, "volume" refers to its mathematical meaning, "space of three or more dimensions," rather than the "volume scan" commonly used in radar research.

Furthermore, by mapping the data volume directly to each rain gauge, the point, we can establish location-specific relationships sensitive to the topographic characteristics. Hence, we called this data-driven approach "volume-to-point" in contrast to the Z-R relationship, which is of point to point nature. Figure 1 illustrated the difference between the point-to-point and volume-to-point approach.

The complexity of the proposed volume-to-point framework is two-fold. Because the aggregated data volume may contain more information than the precipitation at one particular location, finding a proper processing scheme to extract significant features is essential. The second fold of complexity comes from the relationship between the data volume and the precipitation intensity by the rain gauges. Marshall and Palmer (1948) had shown that this relationship should be in the form of the power-law and hence implied that we could start from nonlinear regressors. The studies of machine learning focus on searching patterns in data and offer an efficient approach to tackle our complexity problem. Given a data volume and corresponding values of precipitation, supervised machine learning models can optimize the predictive performance of the unknown relationship (Bishop., 2006; Hastie et al., 2009). In addition to the conceptual volume-to-point framework, we also proposed one implementation based on state-of-the-art machine learning techniques. The feasibility and effectiveness of the proposed method were evaluated based on observation data during 2013 ~ 2016 over the Taipei basin.

The following section describes details of the data used in this study, as well as in-depth elaborations of the implementation of the volume-to-point approach. The evaluation results are shown in section 3, and we will give concluding remarks of the proposed method in section 4.

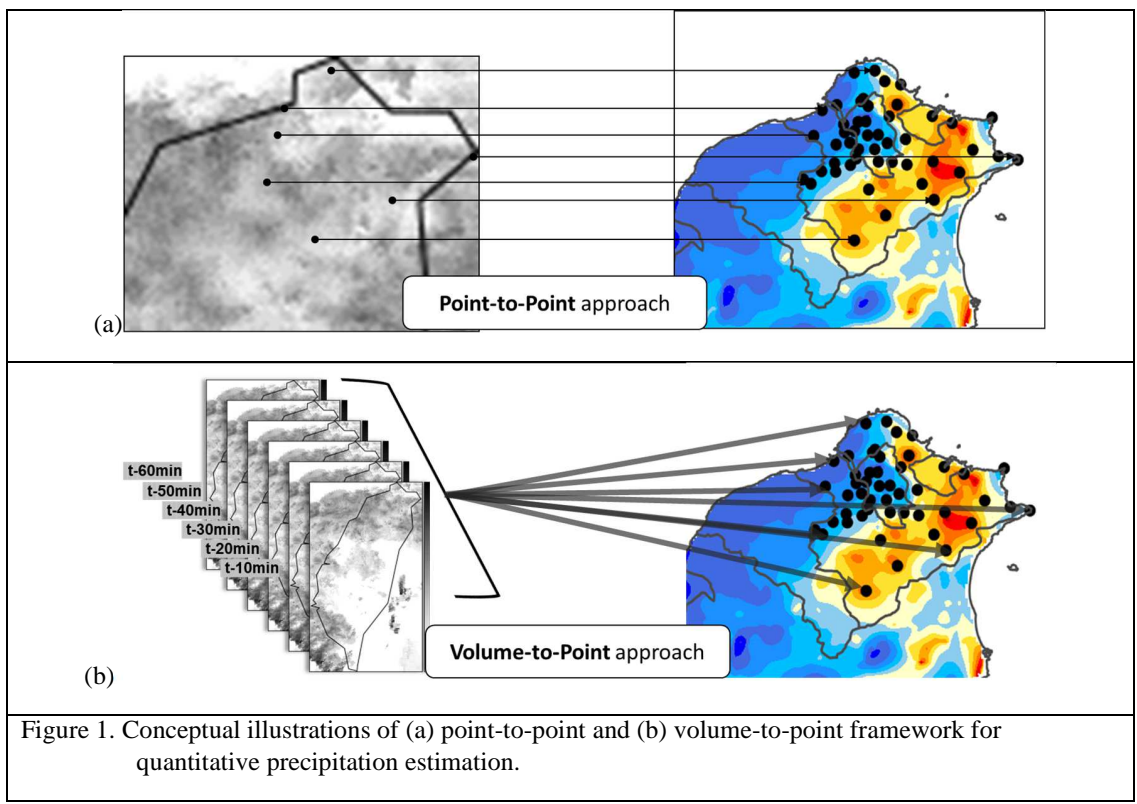

Figure 1. Conceptual illustrations of (a) point-to-point and (b) volume-to-point framework for quantitative precipitation estimation.

**2 Methods and Data**

2.1 Data

In this study, we used two observation data sets from the year 2013 to 2016. QPESUMS Mosaic dataset is a product developed by the Central Weather Bureau (CWB) and the National Severe Storm Laboratory (Gourley et al., 2002; Chiou et al., 2004). It features multiple radar integration with outputting the maximum reflectivity on two-dimensional grids. The product provided by CWB has resolutions of 0.0125°×0.0125° within the area of 21.8875° ~ 25.3125°N / 120.0000° ~ 122.0125°E, and the update interval is 10 minutes. More specifically, for each hour, the dimension of the data volume is 6 x 162 x 275. Figure 2 demonstrated one snapshot of the data volume at 2014-06-14 19:10~20:00 (LST) during typhoon Hagibi. As for the point target data, we used the hourly precipitation from 45 CWB weather stations over the Taipei area as the ground truth. Figure 3 shows the locations of the 45 stations. The weather stations are in the metropolitan area on the low elevation (around 20m above mean sea level) as well as over the surrounding hills (no more than 1000 m above mean sea level) of the Taipei basin. The Feitsui reservoir, the main water supply to millions of residents, is located at the hill to the south of the Taipei basin. Watershed areas are highlighted in blue in figure 3, and the large area in the center represents the Feitsui reservoir watershed region.

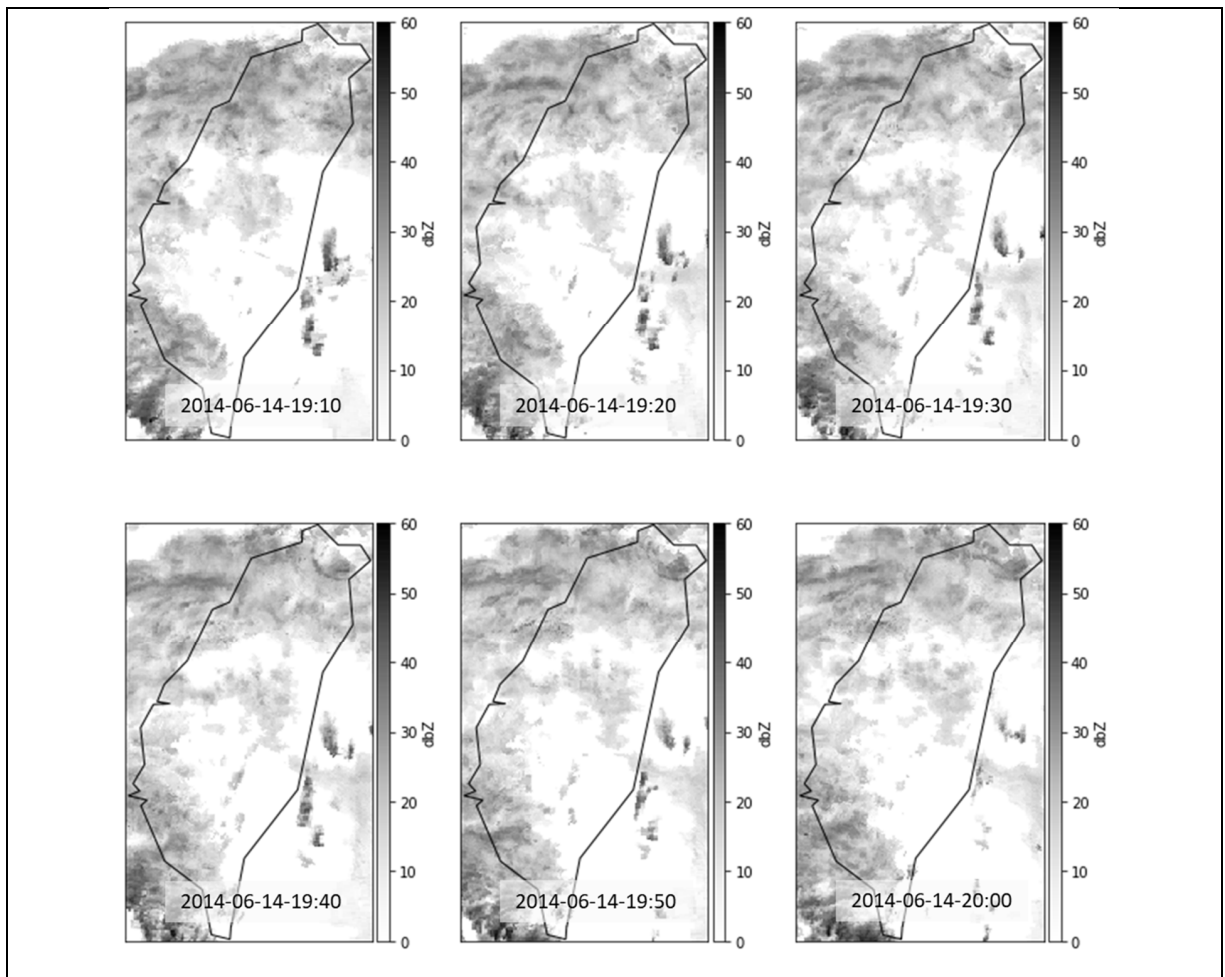

Figure 2. The QPESUMS Mosaic data of 2014-06-14 19:10 ~ 20:00 (UTC+8). At this moment, Taiwan was under the influence of typhoon Hagibi.

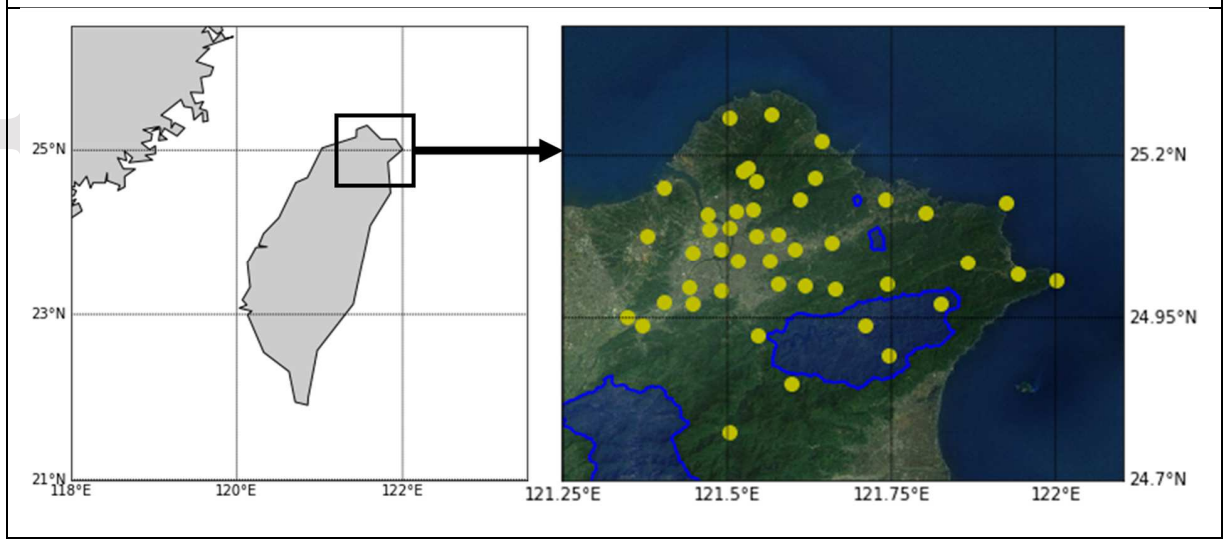

Figure 3. Locations of 45 CWB weather stations over the greater Taipei metropolitan area. The blue shaded areas represent the watershed regions.

2.2 The Volume-to-Point Framework

Conceptually, the volume-to-point framework aims to establish the relationship between a data volume and the corresponding precipitation at a specific location. In this study, we explicitly define the point data as the hourly precipitation of each CWB station, and the volume data as the six 10-min snapshot QPESUMS image before the precipitation record. For example, for the rain gauge measure of one station at 1200Z, the corresponding data volume is the collection of QPESUMS data at 1100Z, 1110Z, 1120Z, 1130Z, 1140Z, and 1150Z.

With this setup, the precise objective of the proposed QPE scheme was to map a data volume of size 6 x 162 x 275 to 1 precipitation value. Mathematically, this is a regression problem, where we take the input of 267,300 independent variables, X, and look for an optimal function, f(X), that yields the best fit to the precipitation, Y. Although fitting a quarter-million independent variables with linear regression is feasible for modern computers, this problem requires more sophisticated treatments. Marshall and Palmer (1948) have shown that the relationship between radar reflectivity and precipitation follows a power law. First, the relationship between spatial-temporal variability and local precipitation is highly nonlinear and hence requires a nonlinear mapping. Second, the main reason we use a large data volume is to attempt to cover the evolution and movement of convective over hills. The cloud dynamics in the presence of hill upslope dynamical forcing from the wind as well as the merger of cloud cells are essential for the heavy precipitation process in the Taipei basin (e.g., Kuo & Wu, 2019; and Miao & Yang, 2020). To fulfill this goal, we need algorithms that can extract spatial and temporal characteristics from the data volume rather than assuming each grid point is independent. Therefore, statistical models that can perform representation learning and nonlinear curve fitting are more suitable for our task.

Machine learning models offer various approaches to regression problems, from simple linear regression to decision trees, kernel methods, and neural networks (Hastie et al., 2009). While simple models generalize better, complicated models can capture nonlinear relationships with the risk of "overfitting." The recent development in deep neural networks has shown significant improvement in classification and regression tasks. By stacking multiple hidden layers in the neural networks, the deep learning methods can discover intricate structures in high-dimensional data (LeCun et al., 2015). Among several variations of deep neural networks, the convolutional neural networks (CNNs) were designed to process data that come in the form of multiple arrays, which is particularly suitable for the QPESUMS data volume.

Figure 4 shows the volume-to-point framework used in this study. As illustrated in figure 4, we use a machine learning model to extract features from the data volume and then map these features to the precipitation of the given weather station. Conventional machine learning methods divided the whole tasks into two processes: feature extraction and regression. Extracting meaningful features from the dataset used to rely on domain knowledge and repetitive try-and-error. Since LeCun first proposed CNN in 1990 (LeCun et al., 1990), CNN-based methods have been applied with great success to various tasks in image processing and computer vision. The strength of CNNs mainly came from their fundamental elements, the convolutional filters (or sometimes referred to as kernels). These convolutional filters are small two-dimensional arrays that scan through the original matrix and then output a new one. With convolutional kernels, the images are processed to preserve local geometric information, rather than treating each point in the image as an independent variable. The use of convolutional kernels is similar to applying two-dimensional operators, such as divergence and curl, to vector fields. And this feature makes CNNs potentially suitable for meteorological datasets. For example, Han and colleagues used CNNs for convective storm nowcasting and showed descent improvement (Han et al., 2020). Shi et al. (2015) and Kalchbrenner & Sønderby (2020) demonstrated using sequential versions of CNNs (ConvLSTM) for precipitation nowcasting.

The convolutional neural networks yielded state-of-the-art performance in computer vision tasks such as image classification and object detection. Unlike traditional neural networks, CNNs employed convolutional kernels and other techniques such as pooling and dropout to form the layers of the network. The convolutional layers consist of multiple kernels that scan through the data matrix input from the previous layer. Max-pooling layers are designed to reduce the size of the preceding data matrix by a factor of k by keeping the maximal value of every k-by-k sub-matrix. The dropout layers randomly select the parameters to be ignored in one training trial and serve as a regularization technique to prevent overfitting in CNNs. Dhillon and Verma (2019) reviewed the research milestones and compared several significant variations of CNNs. The convolutional neural network used in this study was based on the architecture suggested by Simonyan and Zisserman with minor modifications (Simonyan & Zisserman, 2015; Reichstein et al., 2019). The detailed structure of our implantation is shown in figure 5.

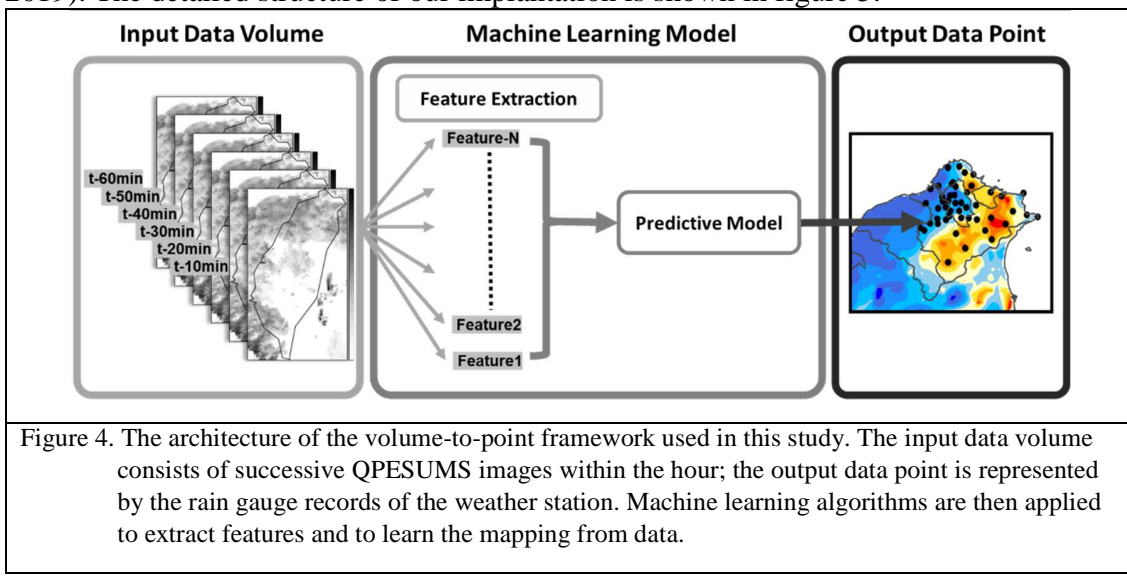

Figure 4. The architecture of the volume-to-point framework used in this study. The input data volume consists of successive QPESUMS images within the hour; the output data point is represented by the rain gauge records of the weather station. Machine learning algorithms are then applied to extract features and to learn the mapping from data.

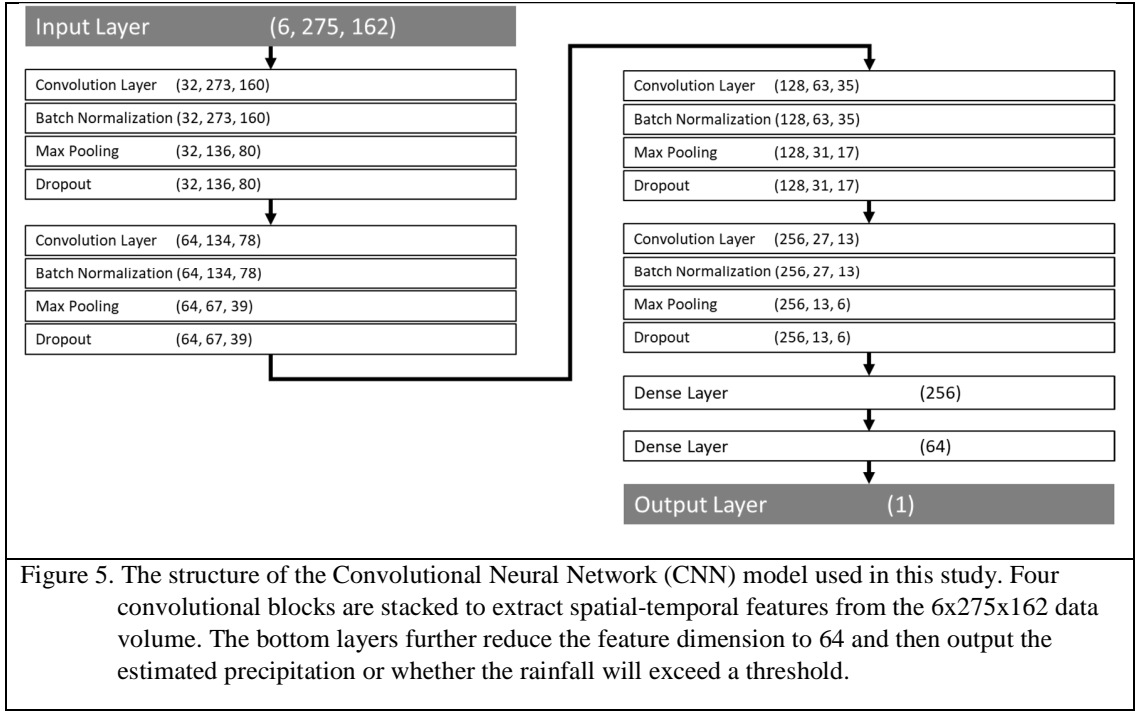

Figure 5. The structure of the Convolutional Neural Network (CNN) model used in this study. Four convolutional blocks are stacked to extract spatial-temporal features from the 6x275x162 data volume. The bottom layers further reduce the feature dimension to 64 and then output the estimated precipitation or whether the rainfall will exceed a threshold.

2.3 Evaluation methods

The volume-to-point framework is based on statistical models, and hence we need to split the data into the training set and the testing set to evaluate the proposed methods. In our experiments, we used data from 2013 to 2015 as the training data and then evaluated the trained models against the data of 2016.

In this study, we evaluated the performance of the volume-to-point framework in two different aspects, i.e., the general errors in estimated precipitation and the ability to detect heavy rainfall. For general performance, we used the RMSE between the measured and predicted rain as the metrics. For the heavy rainfall cases, we used the hit-rate of correctly estimated rain larger than 30 mm/hr as the primary indicator. Also, the same metrics of the CWB operational scheme (based on Z-R relation) were derived for comparison.

Besides the general performance, we also wanted to demonstrate the effect of spatial location and weather events. Thus, the performance metrics at different altitudes and locations, as well as the case of typhoon Megi (2016-09-25 ~ 2016-09-28), were presented in the next section.

3 Results

The proposed volume-to-point approach was trained with the data between 2013 and 2015 and then evaluated with the data in 2016. The averaged RMSE of the proposed volume-to-point method of the 45 stations is 1.86 mm/hr, while the same metric of the CWB operational QPE scheme is 1.90 mm/hr. In general, the proposed approach performed slightly better (i.e., with smaller RMSE), but the difference is not significant (p=0.67 for paired t-test). Figure 6 illustrated the RMSE of two methods for each of the 45 stations arranged in ascending elevation. As shown in figure 6, the performance gain of the proposed volume-to-point framework was more evident at low and high altitudes.

Figure 7 showed the RMSE on the map. The spatial distributions of RMSE are similar for both QPE methods (figure 7 (a) and (b)). The southeastern part of the map is the central mountain ridge that blocked most radar signals, and hence both methods showed higher RMSE in that region. For better comparison, figure 7(c) showed the difference of RMSE between two QPE schemes, while positive values indicate that the volume-to-point approach performed better. In figure 7(c), we can see the proposed framework showed advantages in two areas: the center of the Taipei basin (at the center of the map with dense dots) and the northern part of the southeastern mountains, which is the watershed area of Feitsui reservoir (shaded on the map). These two regions are located next to one of the radar networks, Wufenshan radar (RCWF), and benefited from better data quality.

Besides estimating the amount of precipitation, the volume-to-point framework can also leverage the classification form of machine learning models to detect heavy-rainfall events. The Central Weather Bureau officially defined the heavy rainfall over the Taiwan area as precipitations reach 40mm/hr or 80mm/day. However, in 2016, 5 of the 45 stations never recorded hourly rainfall larger than 40mm. Therefore, we used 30mm/hr, close to the 99.5th percentile of 1960 – 2010 (Su et al., 2012), as the threshold of heavy-rainfall events in this study. Figure 8 showed the hit-rate of the heavy-rainfall events, i.e., the probabilities that QPE methods did report heavy-rainfall when it occurred, of the CWB operational QPE scheme and the volume-to-point approach. As shown in the figure, the proposed method has higher hit-rates for most stations (39 out of 45) with an average of 0.8, in contrast to conventional QPE's averaged hit-rate of 0.34. The high hit-rate came with a price of a high false-alarm rate of 0.0134 on average compared to 0.0005 for the CWB's method. Figure 9 illustrated the receiver operating characteristic of two QPE schemes, where we can see the gain in hit-rate exceeding the loss in false-alarm-rate. Noted that the statistics are derived from each hour of the entire 2016, while the probability of heavy-rainfall events was less than 0.1%. With the low prevalence in nature, a low false-alarm-rate can be expected for any classification algorithm.

Besides the general statistics, figure 10 illustrated the QPE results during the typhoon Megi (2016-09-25 ~ 2016-09-28) at weather station 466920 (Taipei) and 466930 (Zhuzihu). As shown in figure 10, the volume-to-point framework can detect high precipitation more accurately while overestimating the rain trace. Also, the CWB operational scheme tended to underestimate the amount of rainfall in most cases.

From the results above, the volume-to-point method is shown to perform equally well as the state-of-the-art QPE method in general situations, and its advantages lied in heavy-rainfall scenarios. In the next section, we will discuss the robustness and possible extensions of the proposed framework.

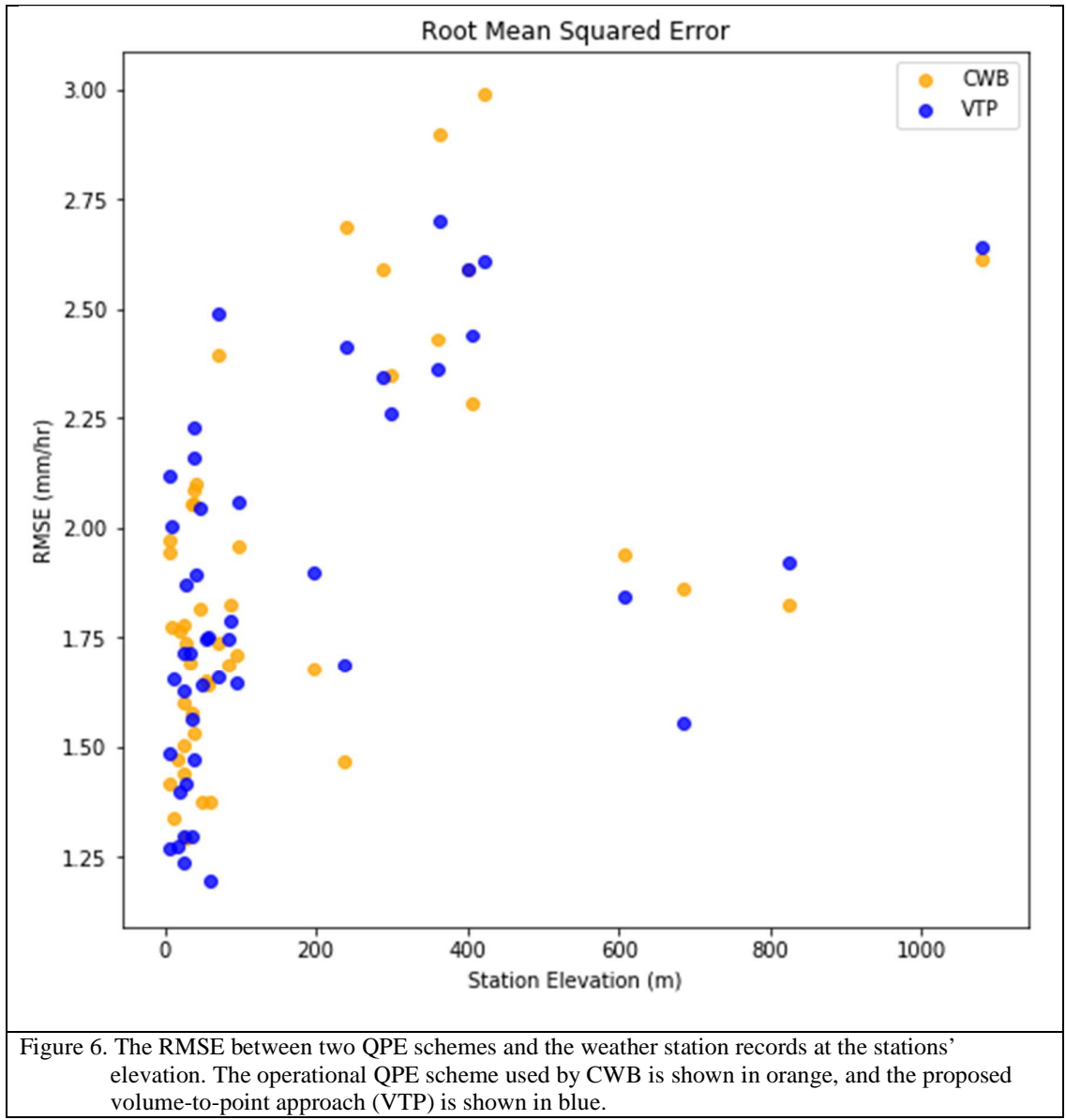

Figure 6. The RMSE between two QPE schemes and the weather station records at the stations' elevation. The operational QPE scheme used by CWB is shown in orange, and the proposed volume-to-point approach (VTP) is shown in blue.

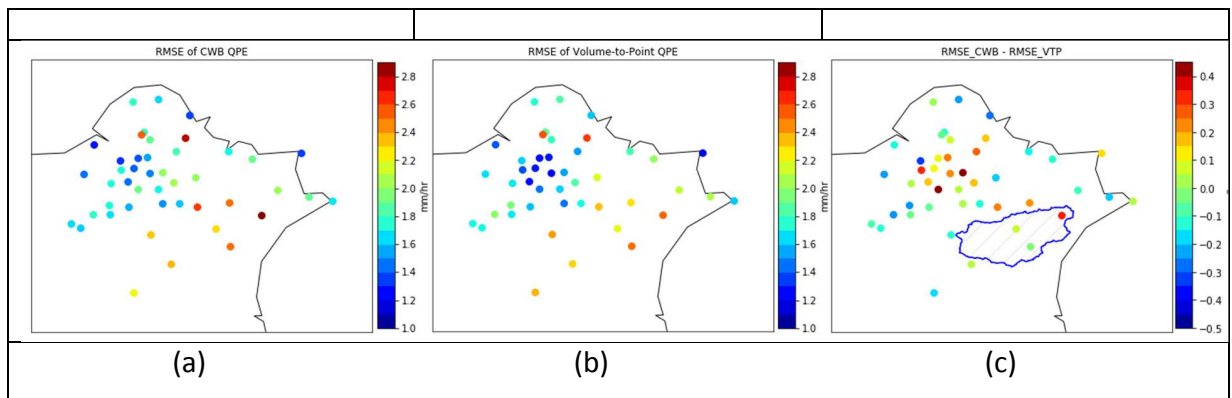

(a) (b) (c)

Figure 7. The RMSE between QPE schemes and the weather station records. The operational QPE scheme used by CWB is shown in the left panel (a), and the proposed VTP approach is shown in the middle (b). The right panel (c) illustrates the RMSE difference, where positive values represent locations the VTP method performed better. The shaded area in (c) represents the watershed regions of Feitsui reservoir.

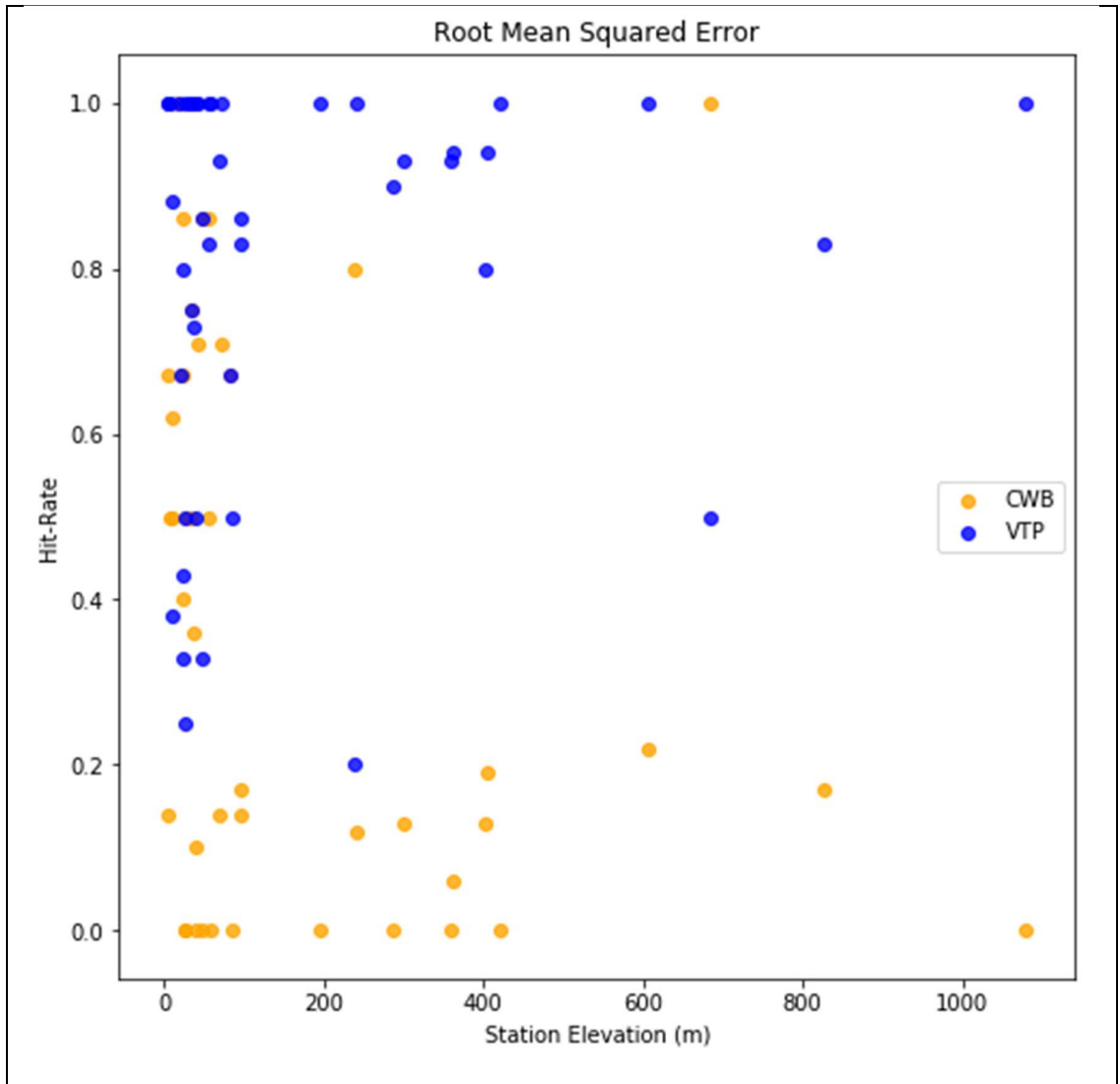

Figure 8. The hit-rate of detecting events of 30mm/hr rainfall of 45 CWB stations and the stations' altitude. The operational QPE scheme used by CWB is shown in orange, and the proposed VTP approach is shown in blue. Stations are arranged in ascending altitude.

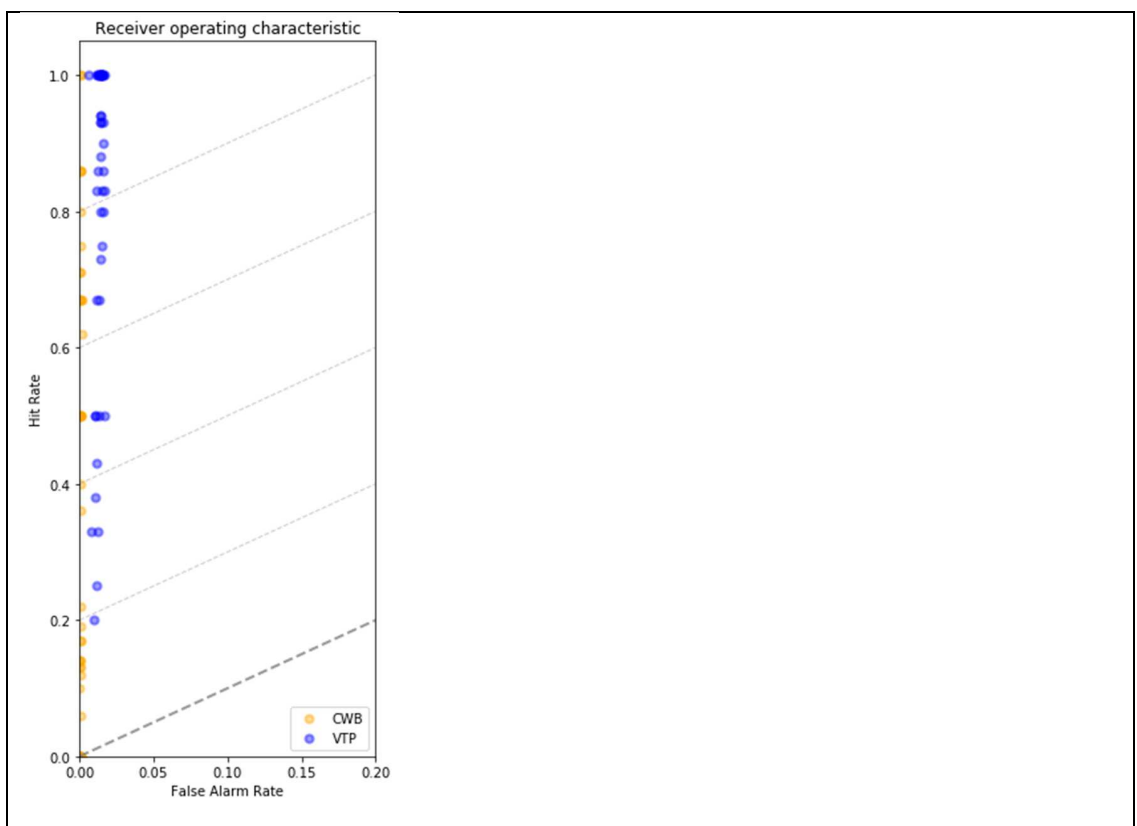

Figure 9. The receiver operating characteristics of two QPE schemes. The CWB operational scheme is shown in orange, and the VTP method is in blue. The dashed lines represent the diagonal, and the points locate more to the upper and right represents higher prediction skill.

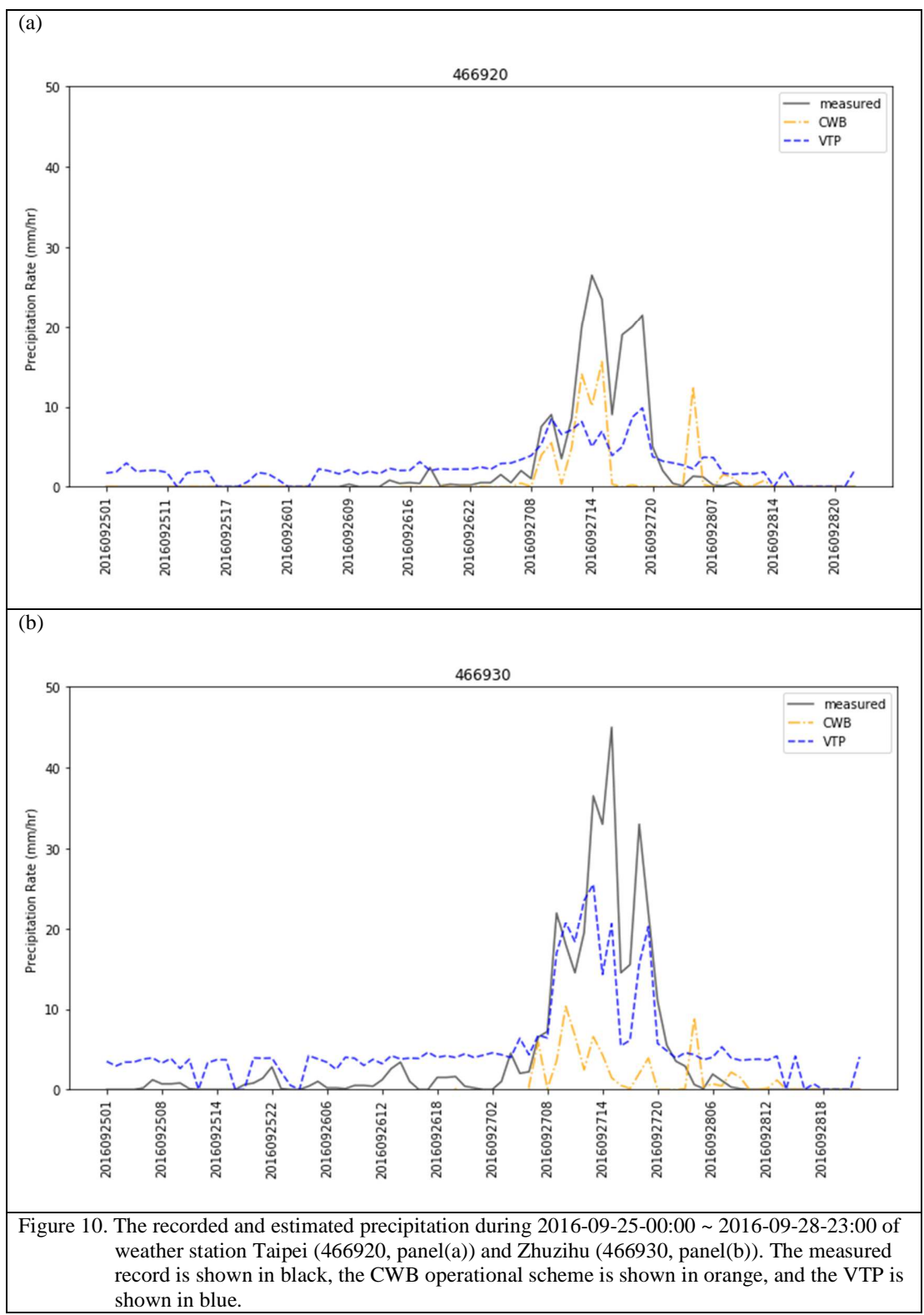

Figure 10. The recorded and estimated precipitation during 2016-09-25-00:00 ~ 2016-09-28-23:00 of weather station Taipei (466920, panel(a)) and Zhuzihu (466930, panel(b)). The measured record is shown in black, the CWB operational scheme is shown in orange, and the VTP is shown in blue.

# 4 Concluding Remarks

### 4.1 Sensitivity of the implementation

The volume-to-point framework used in this study is a data-driven approach. One common obstacle data-driven methods often encounter is generalizability. That is to say, is the performance stable in various situations, or is it only specific for the dataset used for evaluation? In figure 6, we can see the variations of RMSE are similar between the two QPE methods. The standard deviations of RMSE among 45 stations were 0.43 and 0.42 for the CWB scheme and volume-to-point, respectively. This negligible difference suggested that the volume-to-point approach is as robust as Z-R based methods across locations. The results reported in the previous section were based on models trained with QPESUMS data between 2013 ~ 2015 and tested against 2016. We tested a different training-testing data split of 2013 ~ 2014 against 2015, and the results were similar (the mean and standard deviation of RMSEs among 45 stations are 1.98 and 0.37, respectively). The test suggested that the volume-to-point framework is also robust across time.

Even though this framework is robust, unlike the Z-R relationship, the model trained with the data of a particular rain gauge cannot be applied to another weather station. This fact may limit this method to only locations with historical records of precipitation, while the Z-R formula can be used on all grid points with radar signals.

### 4.2 Extending the volume-to-point framework

The method proposed in this study is a framework based on machine learning models, and hence it inherited certain advantages.

First of all, the field of machine learning progressed rapidly in recent years. Though the implementation used in this study was based on the state-of-the-art model of convolutional neural networks, it is very likely future advancements in machine learning can offer better models for such problems. In that case, we can easily replace the feature extraction or the machine learning model shown in figure 4 with more advanced methods without changing the whole procedure.

Second, though our implementation used only the QPESUMS dataset as the input data source, the proposed framework doesn't limit itself to homogeneous data input. For example, the advancement of radar technology can offer much more information than reflectivity. Also, remote sensing data can be integrated as part of the data volume and be processed with the same procedure.

The third advantage is that it is straightforward to transform this framework from quantitative precipitation estimation to forecasting. Machine learning models are mathematical mappings between the input and the output data. In terms of QPE, the input is the radar reflectivity, and the output is the corresponding precipitation at the same time. If we change the output to the rain in the future, then the same framework will represent the forecasting. Figure 11 illustrated the RMSE of using the same one-hour QPESUMS data and machine learning methods for QPF. We selected five weather stations with human crews for more precise visualization. As shown in figure 11, the errors increased rapidly after 1 or 2 hours. The averaged correlation coefficients of one hour QPF is 0.3, and it dropped to 0.07 for 4-hour QPF, which indicated that the QPF results lose its predictive power in time quickly. This result was not surprising because we only use one hour of QPESUMS data for prediction. To improve the QPF, a more extended period of radar observations or other information sources such as the numerical model output of thermodynamic and dynamic variables could be beneficial. Specifically, the temperature profile of convective instability and the three-dimensional wind of cloud dynamics can be incorporated into the proposed method. The wind field information may be important in the problem we have that target area covered some in terrain area. On the other hand, the proposed method with the current setting can produce results compare favorably with state of the art Z-R method with some additional advantages in certain areas. With the results as the reference benchmark, the proposed volume-to-point framework can integrate heterogeneous data sources, as discussed in the earlier paragraph, to improve the QPF forecast.

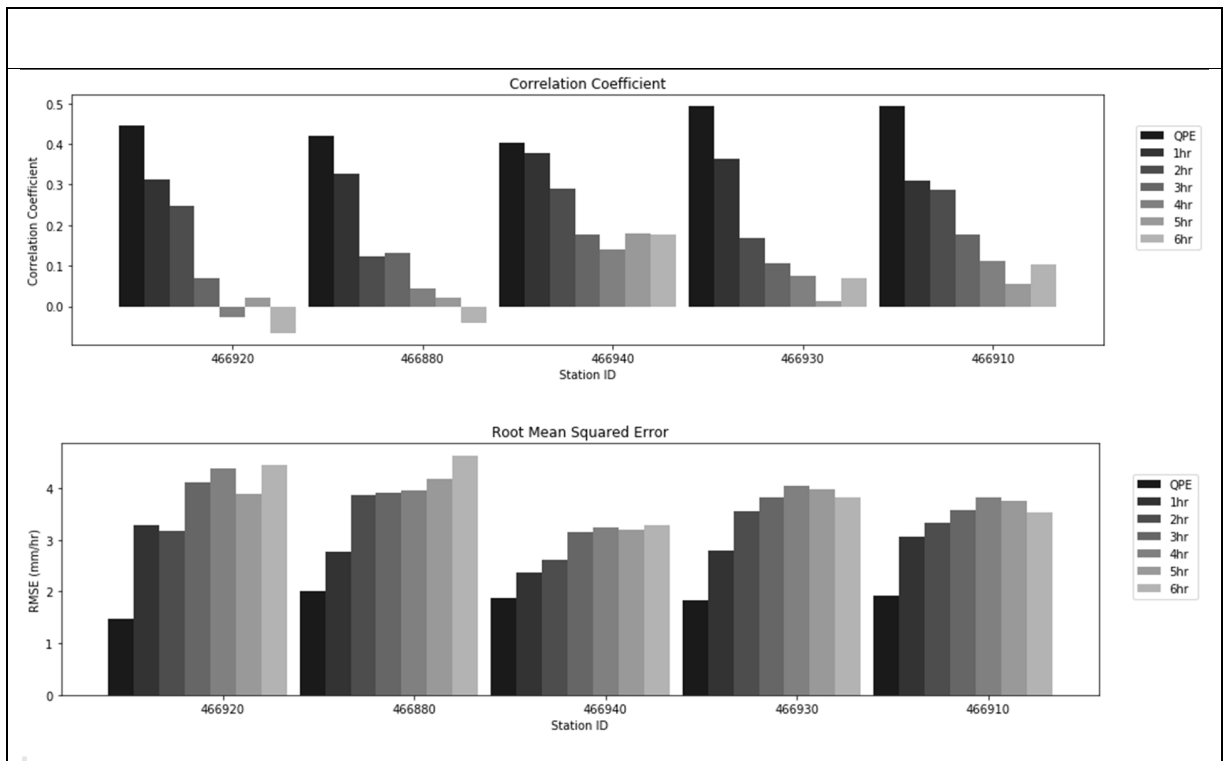

Figure 11. The results of using the same VTP implementation for hourly rainfall forecasting. Only five stations are tested, and they are arranged in ascending altitude. The correlation coefficient dropped rapidly in the first 3 hours, and the RMSE reached peaks in 3~4 hours for all stations. The results suggested we should not infer the future precipitation for more than three hours by using one-hour QPESUMS data.

4.3 Summary

In this study, we introduced the volume-to-point framework for quantitative precipitation estimation from radar reflectivity. Instead of making inferences based on the signal of a single point, our approach used a data volume covering a larger spatial area and a more extended period. The aggregated data volume can provide information about the existence and movement of weather systems, and it is more likely to include cloud dynamics associated with the precipitation process (Miao & Yang, 2020). The convolutional neural networks were chosen to implement the volume-to-point framework to extract essential features from a massive dataset. We trained the model with data in 2013~2015 and evaluated the QPE performance in 2016.

In comparison to the operational QPE scheme used by the Central Weather Bureau, the volume-to-point framework performed comparably well in general cases and excelled in detecting heavy-rainfall events. In terms of geographical locations, our method performed particularly well in the highly populated area and the major watershed (figure 7c), which is significant in practice. These two regions locate next to the primary radar in Taipei, and hence we infer the proposed method may directly benefit from good data quality.

Due to the limitation of the chosen CNN architecture, the features learned by the models were difficult to visualize, as suggested in McGovern et al. (2019). However, the presented results indicated that providing a more extensive data volume and using a specific model for each rain gauge did improve the ability to detect heavy-rainfall events.

Finally, we want to emphasize that the presented work can serve as a baseline of the volume-to-point framework for future development. As discussed in the previous section, the proposed framework can be further extended to conduct precipitation forecasts and to incorporate heterogeneous data sources. Our preliminary test showed that with a data volume consists of only one-hour radar reflectivity, our approach can perform very well as a QPE scheme and can provide a reference in forecasting. As on-going work, we will further explore the feasibility of integrating different data types such as numerical model wind field and temperature profile and use the volume-to-point framework as a QPF scheme in extreme precipitation scenarios.

### Acknowledgments

This research was supported by the Taiwan Ministry of Science and Technology through grants MOST-107-2111-M-034-003, MOST-108-2119-M-002-022, and MOST-108-2625-M-034-002. We would also like to show our gratitude to the Central Weather Bureau and Data Bank for Atmospheric and Hydrologic Research for providing the raw QPESUMS data and the hourly precipitation data.

### Data Availability Statement

The hourly precipitation data of 45 weather stations can be downloaded from the Open Data Platform (OSF) at https://osf.io/pkxu6/. The features extracted from the QPESUMS data volume and the estimated precipitations of the 45 weather stations can be downloaded from our OSF repository, along with the codes creating figures reported in this study.

**Figure 1**. Conceptual illustrations of (a) point-to-point and (b) volume-to-point framework for quantitative precipitation estimation.

**Figure 2.** The QPESUMS Mosaic data of 2014-06-14 19:10 ~ 20:00 (UTC+8). At this moment, Taiwan was under the influence of typhoon Hagibi.

**Figure 3.** Locations of 45 CWB weather stations over the greater Taipei metropolitan area. The blue shaded areas represent the watershed regions.

**Figure 4.** The architecture of the volume-to-point framework used in this study. The input data volume consists of successive QPESUMS images within the hour; the output data point is represented by the rain gauge records of the weather station. Machine learning algorithms are then applied to extract features and to learn the mapping from data.

**Figure 5.** The structure of the Convolutional Neural Network (CNN) model used in this study. Four convolutional blocks are stacked to extract spatial-temporal features from the 6x275x162 data volume. The bottom layers further reduce the feature dimension to 64 and then output the estimated precipitation or whether the rainfall will exceed a threshold.

**Figure 6.** The RMSE between two QPE schemes and the weather station records at the stations' elevation. The operational QPE scheme used by CWB is shown in orange, and the proposed volume-to-point approach (VTP) is shown in blue.

**Figure 7.** The RMSE between QPE schemes and the weather station records. The operational QPE scheme used by CWB is shown in the left panel (a), and the proposed VTP approach is shown in the middle (b). The right panel (c) illustrates the RMSE difference, where positive values represent locations the VTP method performed better. The shaded area in (c) represents the watershed regions of Feitsui reservoir.

**Figure 8.** The hit-rate of detecting events of 30mm/hr rainfall of 45 CWB stations and the stations' altitude. The operational QPE scheme used by CWB is shown in orange, and the proposed VTP approach is shown in blue. Stations are arranged in ascending altitude.

**Figure 9.** The receiver operating characteristics of two QPE schemes. The CWB operational scheme is shown in orange, and the VTP method is in blue. The dashed lines represent the diagonal, and the points locate more to the upper and right represents higher prediction skill.

**Figure 10.** The recorded and estimated precipitation during 2016-09-25-00:00 ~ 2016-09-28-23:00 of weather station Taipei (466920, panel(a)) and Zhuzihu (466930, panel(b)). The measured record is shown in black, the CWB operational scheme is shown in orange, and the VTP is shown in blue.

**Figure 11.** The results of using the same VTP implementation for hourly rainfall forecasting. Only five stations are tested, and they are arranged in ascending altitude. The correlation coefficient dropped rapidly in the first 3 hours, and the RMSE reached peaks in 3~4 hours for all stations. The results suggested we should not infer the future precipitation for more than three hours by using one-hour QPESUMS data.